\documentclass[prb,twocolumn,showpacs,amsmath,amssymb,floatfix]{revtex4}
\usepackage{graphicx}

\usepackage{epstopdf}

\def\beq{\begin{equation}}
\def\eeq{\end{equation}}

\newcommand{\qp}{Q_{\textrm{P}}}
\newcommand{\qir}{Q_{\textrm{IR}}}
\newcommand{\omp}{\Omega_{\textrm{P}}}
\newcommand{\omir}{\Omega_{\textrm{IR}}}

\begin{document}

\title{Proposal for ultrafast switching of ferroelectrics using mid-infrared pulses}

\author{Alaska Subedi} 

\affiliation{Max Planck Institute for the Structure and Dynamics of
  Matter, Luruper Chaussee 149, 22761 Hamburg, Germany}

\date{\today}

\begin{abstract}
  I propose a method for ultrafast switching of ferroelectric
  polarization using mid-infrared pulses. This involves selectively
  exciting the highest frequency $A_1$ phonon mode of a ferroelectric
  material with an intense mid-infrared pulse. Large amplitude
  oscillations of this mode provides a unidirectional force to the
  lattice such that it displaces along the lowest frequency $A_1$
  phonon mode coordinate because of a nonlinear coupling of the type
  $g\qp\qir^2$ between the two modes. First principles calculations
  show that this coupling is large in perovskite transition-metal
  oxide ferroelectrics, and the sign of the coupling is such that the
  lattice displaces in the switching direction. Furthermore, I find
  that the lowest frequency $A_1$ mode has a large $\qp^3$ order
  anharmonicity, which causes a discontinuous switch of electric
  polarization as the pump amplitude is continuously increased.
\end{abstract}


\maketitle

\section{Introduction}

Ultrafast switching of polarization in ferroelectrics is of great
interest for potential application in non-volatile memory
devices. FLASH memories, which at present are the most commonly used
non-volatile memory devices, have an operating speed of milliseconds.
Because of their slow speed, they are not considered as candidates for
future memory applications \cite{jeon12}. Other emerging non-volatile
memory technologies that utilize phase or resistance change have write
and erase times of nanoseconds. Therefore, development of a switching
mechanism at sub-picosecond timescales has the potential to
revolutionize the field. 

Non-destructive readout of the electric polarization at sub-picosecond
timescales has recently been demonstrated by analyzing the THz pulse
waveforms radiated after illumination of a ferroelectric sample by
femtosecond laser pulses at optical wavelengths \cite{taka06,
  talb08}. This makes ferroelectric materials an exciting prospect for
memory applications if switching can be achieved at similarly
ultrashort timescales.

Ferroelectric materials exhibit remnant polarization even at zero
external electric field because the cations and anions in these
materials are asymmetrically displaced in the equilibrium structure.
To switch the polarization, the relative displacement between the
cations and anions need to be reversed. This can be achieved by
applying a (quasi-)static electric field because such an electric
field imparts a unidirectional force to the cations and anions. An
arbitrary light pulse, whose oscillating electric field integrates to
zero by definition, imparts a zero total force to the electric dipole
present in the material. Therefore, an ultrashort light pulse cannot
in general be used to switch the polarization of a ferroelectric
material. Nevertheless, there have been several proposals for
switching the polarization of ferroelectric materials using ultrashort
light pulses by controlling their soft phonon modes\cite{fahy94,qi09}.

In this paper, I propose a method for using mid-infrared pulses that
are resonant with the highest frequency infrared-active phonon mode of
a perovskite transition-metal oxide ferroelectric to switch its
polarization. This involves controlling the dynamical degrees of
freedom of the lattice and requires four main ingredients. First, I
notice that there always exists a low frequency fully symmetric $A_1$
phonon mode in the ferroelectric phase that involves the motion of the
cations and anions of the material in a way that changes the electric
polarization. Second, I find that this phonon mode couples to the
highest frequency infrared-active $A_1$ phonon mode of the material
with a large $g \qp \qir^2$ coupling, where $g$ is the coupling
constant and $\qp$ and $\qir$ are the normal mode coordinates of the
lowest frequency and highest frequency $A_1$ normal mode coordinates,
respectively. Third, I find from first principles calculations that
the sign of the coupling is such that the excitation of the highest
frequency $A_1$ mode provides a displacive force along the $\qp$
normal mode coordinate in the direction that switches the
polarization. Fourth, I find that the $\qp$ mode has a strong $\qp^3$
order anharmonicity, which facilitates an abrupt switch of electric
polarization as the $\qir$ amplitude is continuously increased.
Coherent displacement along Raman mode coordinates utilizing nonlinear
phonon couplings by resonantly exiting the highest frequency infrared
mode of various centrosymmetric oxides has previously been
demonstrated \cite{fors11,fors13,sube14,mank14}. Therefore, the method
proposed here to switch ferroelectric polarization at ultrafast
timescales using mid-infrared pulses is experimentally feasible.

\section{Computational details}

I illustrate the proposed mechanism for the case of PbTiO$_3$. The
phonon frequencies and eigenvectors and the nonlinear couplings
between two phonon modes were obtained using density functional theory
calculations with plane-wave basis sets and projector augmented wave
pseudopotentials \cite{bloc94, kres99} as implemented in the {\sc
  vasp} software package \cite{kres96}. The interatomic force
constants were calculated using the frozen-phonon method
\cite{parl97}, and the {\sc phonopy} software package was used to
calculate the phonon frequencies and eigenvectors \cite{togo08}. Total
energy calculations were then performed as a function of the
lowest frequency $\qp$ and high-frequency $\qir$ coordinates to obtain
energy surfaces. The nonlinear couplings between the two modes were
obtained by fitting the calculated energy surface to the polynomial
shown in Eq.~\ref{eq:energy}.

I used the experimental values of $a$ = 3.9039 and $c$ = 4.1348
\AA\ for the tetragonal lattice parameters but relaxed the atomic
positions. The calculations were performed within the local density
approximation. A cut-off of 600 eV was used for the plane-wave basis
set expansion, and an $8 \times 8 \times 8$ $k$-point grid was used in
the Brillouin zone integration.


\section{Results and discussions}

\begin{table}[b]
  \caption{\label{tab:freq} The zone-center phonon frequencies and
    their irreducible representation and optical activity of
    ferroelectric PbTiO$_3$.}
  \begin{ruledtabular}
    \begin{tabular}{lll}
      frequency (cm$^{-1}$) & irrep & optical activity \\
      \hline
      80   & $E$   & IR + Raman \\
      149  & $A_1$ ($\qp$) & IR + Raman \\
      187  & $E$   & IR + Raman \\
      273  & $E$   & IR + Raman \\
      290  & $B_1$ & Raman \\
      356  & $A_1$ & IR + Raman \\
      491  & $E$   & IR + Raman \\
      655  & $A_1$ ($\qir$) & IR + Raman
    \end{tabular}
  \end{ruledtabular}
\end{table}

The ferroelectric phase of PbTiO$_3$ exists in the $P4mm$ structure
with one formula unit per unit cell. This gives rise to 12 zone-center
optical normal modes with the decomposition $\Gamma_{\textrm{optic}} =
3A_1 + B_1 + 4 E$. The $A_1$ and $E$ modes are both Raman and infrared
active, whereas the $B_1$ is only Raman active. The calculated zone
center phonon frequencies and their symmetries are given in Table
\ref{tab:freq}. I find that the coupling between the lowest frequency
$E$ and the highest frequency $A_1$ mode is weak. Therefore, I focus
on the coupled dynamics of the lowest and highest frequency $A_1$
modes.

\begin{figure}
  \includegraphics[width=\columnwidth]{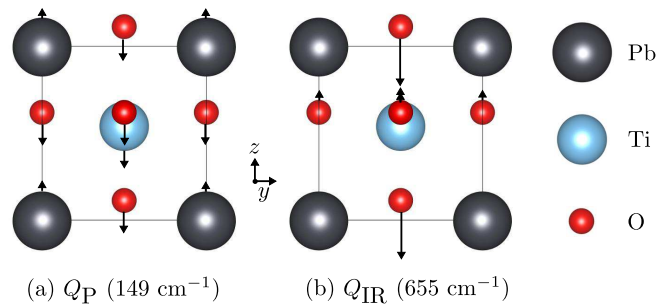}
  \caption{(Color online) Displacement patterns of the (a) lowest
    frequency (149 cm$^{-1}$) $\qp$ and (b) highest frequency (655
    cm$^{-1}$) $\qir$ modes of the ferroelectric phase of
    PbTiO$_3$. Both modes belong to the $A_1$ irreducible
    representation.}
  \label{fig:disp}
\end{figure}

The atomic displacement pattern of the lowest frequency $A_1$ mode
(denoted by $\qp$) is shown in Fig.~\ref{fig:disp}(a) with $z$ axis
chosen as the polarization axis. A finite magnitude of this mode
involves the motion of Pb$^{2+}$ and O$^{2-}$ ions along the $z$ axis
in the opposite direction, and a displacement of the lattice along the
coordinate of this mode modifies the electric polarization of the
material. In the convention used in this paper, a large negative value
of this normal mode coordinate would reverse the polarization. The
arrows in Fig.~\ref{fig:disp}(a) indicate the movements of ions for
such a negative value of $\qp$. One can see that such a movement
reverses the relative displacement between the Pb$^{2+}$ and O$^{2-}$
ions. However, it should be noted that a displacement along this mode
does not bring the structure to the symmetrically equivalent ground
state with opposite polarization because the eigenvector of this mode
is not in general equal to the eigenvector of the unstable infrared
mode of the paraelectric phase that is responsible for the
ferroelectric instability. A further relaxation of the lattice, in
addition to the large negative value of the $\qp$ coordinate that
reverses the polarization, would take the structure to the
symmetrically equivalent switched ground state. I confirmed this by
starting with a structure that was displaced by a value of $-$8 \AA
$\sqrt{\textrm{amu}}$ along the $\qp$ coordinate and relaxing the
atomic positions by minimizing the forces. I found that the structure
indeed relaxes to the symmetrically equivalent switched phase rather
than going back to the initial ferroelectric equilibrium state that
was used as a starting point to displace along the $\qp$ coordinate.

Therefore, a coherent displacement of the lattice along this low
frequency $\qp$ phonon mode coordinate is a viable route for ultrafast
ferroelectric switching. In fact, Qi \textit{et al.} have proposed a
method for switching the polarization by driving large amplitude
oscillations of this phonon mode using multiple THz pulses with an
asymmetric electric field profile \cite{qi09}.

Here I propose a method of switching the polarization using a light
pulse that does not directly drive the low frequency $\qp$
mode. Instead, this involves exciting the high frequency $\qir$
infrared mode of the material (shown in Fig.~\ref{fig:disp}(b)) by an
intense mid-infrared pulse that in turn provides a displacive force
along the $\qp$ coordinate in the switching direction due to a
nonlinear coupling of the type $g\qp\qir^2$ between the two modes.
Furthermore, the presence of a large $\qp^3$ order anharmonicity
causes a sudden increase in the displacement along the $\qp$
coordinate as the $\qir$ amplitude is continuously increased, and this
causes an abrupt reversal of the electric polarization without the
magnitude of the polarization going to zero.

\begin{figure}
  \includegraphics[width=\columnwidth]{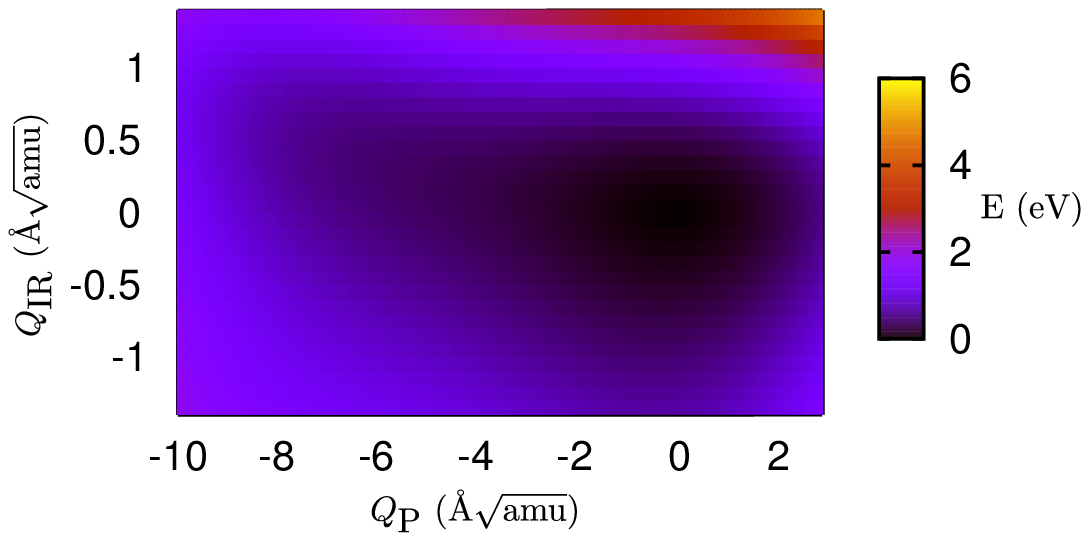}
  \includegraphics[width=\columnwidth]{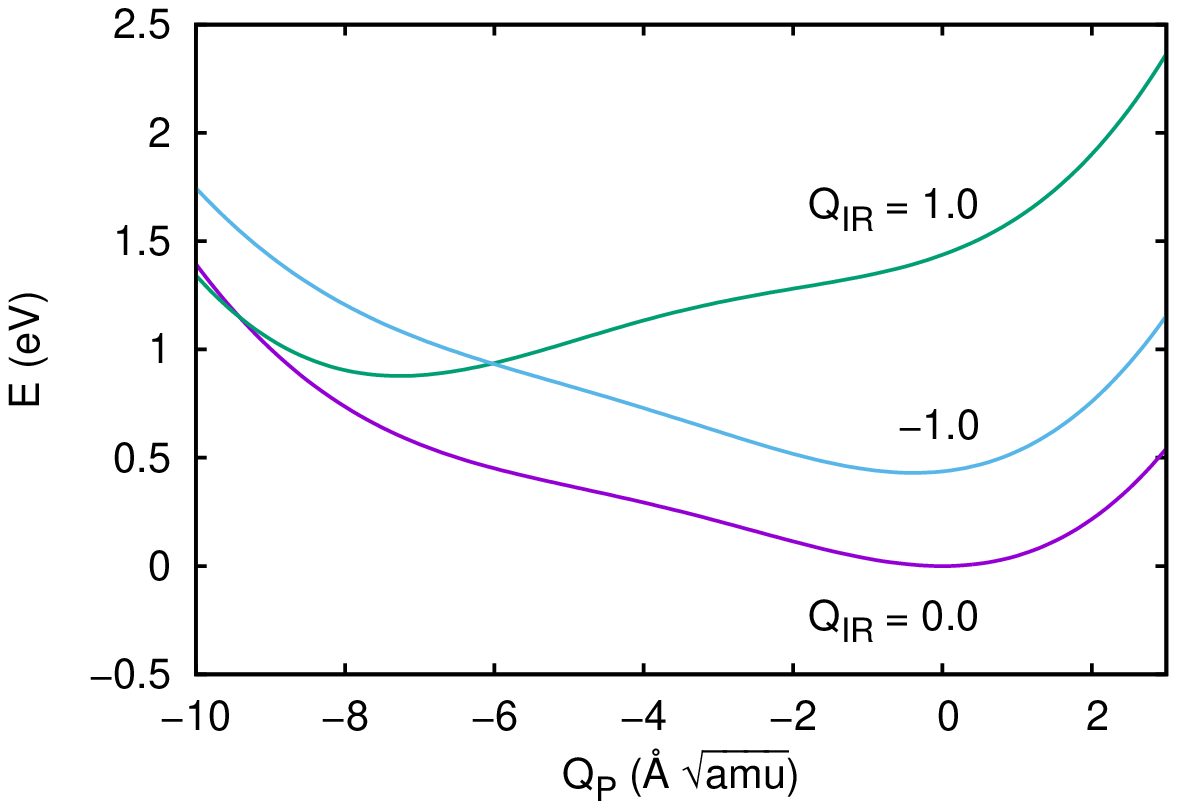}
  \caption{(Color online) Total energy as a function of the $\qp$ and $\qir$
    normal mode coordinates of the ferroelectric PbTiO$_3$. Top: energy
    surface. Bottom: few energy curves that illustrate the behavior of
    the $\qp$ mode as a function of $\qir$ mode.}
  \label{fig:ene}
\end{figure}

I calculated the total energy as a function of the lowest frequency
$\qp$ and highest frequency $\qir$ infrared mode coordinates from
first principles using density functional theory calculations. The
calculated energy surface of ferroelectric PbTiO$_3$ is shown in
Fig.~\ref{fig:ene}, and it fits the following expression:
\begin{eqnarray}
  V(\qp,\qir) & = & \frac{1}{2} \omp^2 \qp^2 + \frac{1}{2} \omir^2
  \qir^2 + \frac{1}{3} a_3 \qp^3 \nonumber \\
  & & + \frac{1}{4} a_4 \qp^4 + \frac{1}{3} b_3 \qir^3 + \frac{1}{4}
  b_4 \qir^4 \nonumber \\
  & & + g\, \qp \qir^2 + h\, \qp^2 \qir + i\, \qp^3 \qir \nonumber \\
  & & + j\, \qp \qir^3 + k\, \qp^2 \qir^3 + l\, \qp \qir^4.
  \label{eq:energy}
\end{eqnarray}
A fit of the above expression to the calculated energy surface
determines \textit{ab initio} the nonlinear couplings between the two
modes up to all significant orders of the two phonon coordinates. The
values of the coefficients of the coupling terms obtained from such a
fit are given in Table II.  The calculated energy surface exhibits
complex features, and this is due to the presence of both even and odd
order nonlinearities. However, there are some salient features.  When
$\qir = 0$, the energy curve of the $\qp$ mode has one minimum at
zero, as one would expect for a stable ground-state structure.  The
energy increases rapidly for positive values of $\qp$, but the
increase is less rapid for negative values of $\qp$. In fact, for
negative values of $\qp$, the slope of the energy curve has a minimum
near $-$5 $\textrm{\AA}\sqrt{\textrm{amu}}$, where the energy curve is
shallow and the restoring force is small, before it shows an upturn
around $-$8 $\textrm{\AA}\sqrt{\textrm{amu}}$. This asymmetric nature
of the energy curve of the $\qp$ mode is due to the presence of a
large $a_3\qp^3$ term in the polynomial expression of the
energy surface. The physical reason for the asymmetric nature of the
$\qp$ energy curve is the presence of a state with reversed
polarization near a $\qp$ value of $-$8
$\textrm{\AA}\sqrt{\textrm{amu}}$ that is symmetrically equivalent to
the ferroelectric state at a $\qp$ value of zero.

\begin{table}[h!tbp]
  \caption{\label{tab:coup} The anharmonic terms and nonlinear
    couplings of the $\qp$ and $\qir$ modes of ferroelectric PbTiO$_3$
    determined from a fit to the energy surface calculated from first
    principles.}
  \begin{ruledtabular}
    \begin{tabular}{lr}
      coefficient & value \, \\
      \hline
      $a_3$ (meV/amu$^{3/2}$/\AA$^3$)   &  21.80    \, \\
      $a_4$ (meV/amu$^{2}$/\AA$^4$)     &  1.89     \, \\
      $b_3$ (meV/amu$^{3/2}$/\AA$^3$)   &  1567.65  \, \\
      $b_4$ (meV/amu$^{2}$/\AA$^4$)     &  631.80   \, \\
      $g$ (meV/amu$^{3/2}$/\AA$^3$)     &  70.32    \, \\
      $h$ (meV/amu$^{3/2}$/\AA$^3$)     &  $-$12.40 \, \\
      $i$ (meV/amu$^{2}$/\AA$^4$)       & $-$0.79   \, \\  
      $j$ (meV/amu$^{2}$/\AA$^4$)       &  52.14    \, \\
      $k$ (meV/amu$^{5/2}$/\AA$^5$)     &  2.29     \, \\
      $l$ (meV/amu$^{5/2}$/\AA$^5$)     &  7.61     \, \\
    \end{tabular}
  \end{ruledtabular}
\end{table}

The energy surface is also asymmetric in the $\qir$ coordinate because
of the presence of odd order terms. The presence of both even and odd
order terms is consistent with the fact that this high frequency mode
also has the $A_1$ representation that does not break any crystal
symmetry and is both infrared and Raman active. Although the energy
surface is asymmetric in the $\qir$ coordinate, both positive and
negative $\qir$ displacements move the minimum of the $\qp$ coordinate
towards the negative direction. A negative value of $\qir$ displaces
the lattice towards the negative $\qp$ direction by a modest amount.
However, a positive value of $\qir$ does not continuously shift the
energy minimum along the negative $\qp$ direction.  It raises the
energy curve near $\qp \approx 0$ and creates an energy minimum at a
large negative value of the $\qp$ coordinate such that a state with
reversed polarization is energetically favored.  This abrupt shift of
the minimum of the $\qp$ mode is due to the presence of a large
$a_3\qp^3$ term.

If the oscillations along the $\qir$ coordinate are integrated out,
the average potential experienced by the lattice has a minimum at a
negative value of the $\qp$ coordinate with the magnitude of the
displacement along $\qp$ coordinate depending on the integration
cut-off (i.e., the amplitude of the $\qir$ oscillations).  This is
because the coupling constant $g$ of the term $\qp \qir^2$ has the
largest magnitude, and it implies that the lattice will experience a
large unidirectional force $-\partial V / \partial \qp = -g \qir^2$
along the $\qp$ coordinate when the $\qir$ mode is being
driven. Furthermore, a large $a_3\qp^3$ term ensures that
the displacement of the lattice along the $\qp$ coordinate abruptly
increases as the $\qir$ amplitude is continuously increased, which
causes the electric polarization to switch discontinuously.

We can achieve a better understanding of the dynamics of the lattice
when the $\qir$ mode is pumped externally by a mid-infrared pulse by
treating the $\qp$ and $\qir$ modes as classical oscillators and
studying their coupled equations of motion. In this picture, the two
oscillators experience a force deriving from the calculated energy
surface, and the $\qir$ mode is additionally driven by a term $F(t) =
F \sin(\Omega t) e^{-t^2/2\sigma^2}$, where $F$, $\sigma$, and
$\Omega$ are the amplitude, width, and frequency of the mid-infrared
pulse, respectively. By treating the expression in Eq.~\ref{eq:energy}
as the potential, the coupled equations motion are
\begin{eqnarray}
  \ddot{Q}_{\textrm{IR}}+\omir^2\qir&=& - b_3 \qir^2 -
  b_4 \qir^3 - 2 g\,\qp\qir \nonumber \\
  & & - h\, \qp^2 - i\, \qp^3 - 3 j\, \qp \qir^2 \nonumber \\
  & & - 3 k\, \qp^2 \qir^3 - 4 l\, \qp \qir^3 +  F(t), \nonumber \\
  \ddot{Q}_{\textrm{P}} + \omp^2 \qp & = & - a_3 \qp^2 - a_4 \qp^3
  - g\, \qir^2 \nonumber \\
  & & - 2 h\, \qp\qir - 3 i\, \qp^2 \qir - j\, \qir^3 \nonumber \\
  & &  - 2 k\, \qp \qir^3 - l \, \qir^4.
\end{eqnarray}

\begin{figure}
  \includegraphics[width=\columnwidth]{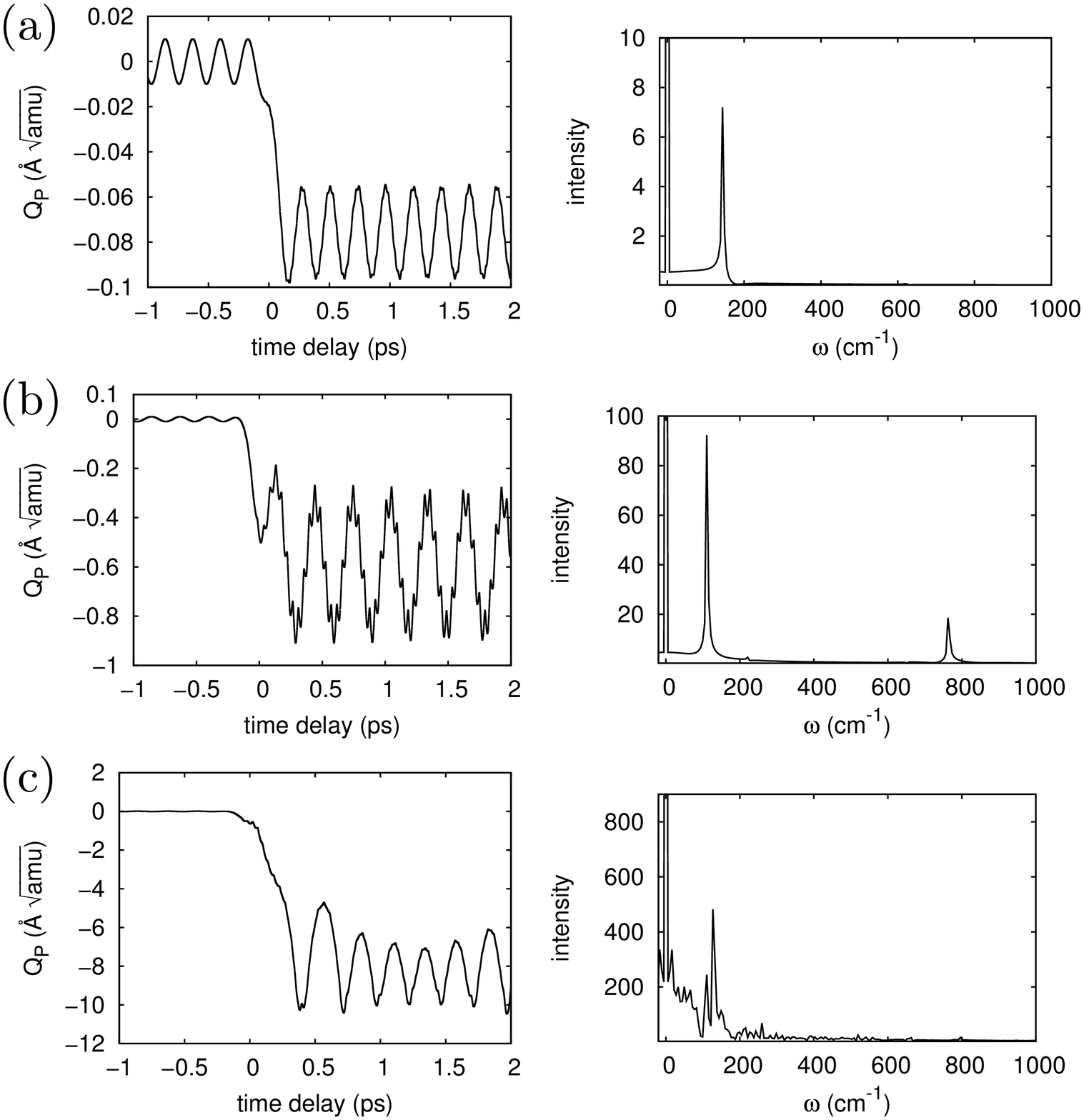}
  \caption{Dynamics of the $\qp$ mode for three different pump
    amplitudes. Left panels: Displacements along $\qp$ coordinate as
    function of time delay. Right panels: Fourier transform of the
    positive time delay oscillations.}
  \label{fig:dyn}
\end{figure}

The results from numerical integration of the coupled equation of
motions for different pump amplitudes are shown in
Fig.~\ref{fig:dyn}. In these calculations, I have used a pump pulse
with a symmetric Gaussian profile and a width of $\sigma = 250$ fs,
which correspond to typical pump pulses used in mid-infrared
excitations \cite{fors11,fors13,mank14}. A pump frequency of $\Omega =
1.03\,\omir$ was used in the simulations, which is chosen to be
slightly off-resonance with the frequency of the $\qir$ mode to
demonstrate that this method is efficacious even if the pump pulse is
not precisely resonant with the $\qir$ mode.

Even for small pump amplitudes that cause a change in the Ti--apical O
distances of a few percent along the $\qir$ coordinate, the $\qp$ mode
oscillates at a displaced position in the negative $\qp$ direction
(see the left panel of Fig.~\ref{fig:dyn}(a)). This is consistent with
the analysis of the $g\qp\qir^2$ coupling presented in
Refs.~\onlinecite{fors11} and \onlinecite{sube14}. In
Ref.~\onlinecite{sube14}, it was shown that the $\qp$ mode experiences
an effective force $-g\qir^2 \propto -gF^2\omir^2\sigma^6(1 - \cos
2\omir t)$ when the $\qir$ mode is pumped by an external driving term
$F(t)$, and the time average of this forcing field has a rectified
non-zero value. The oscillation of the $\qp$ mode at a rectified
position is also seen in the Fourier transform of the time evolution
of the $\qp$ mode at positive time delays as shown in the right panel
of Fig.~\ref{fig:dyn}(a). The Fourier transform shows a peak at zero
frequency that is due to the displacement of the lattice along the
$\qp$ coordinate. In addition, there is a peak at $\omp$ and a
negligible presence of higher-harmonics, which shows that the dynamics
of the coupled oscillators are determined by the $g\qp\qir^2$ term,
and other nonlinearities only play a marginal role at small pump
amplitudes. At small pump amplitudes, the displacement along the
negative direction in the $\qp$ coordinate is small. Therefore,
although the electric polarization is reduced, the reversal of the
polarization has not occurred.

It is also noteworthy that the cubic terms in the energy potential
$a_3\qp^3$ and $b_3\qir^3$ are large. These also
impart unidirectional forces $-\partial V / \partial \qp = -a_3\qp^2$
and $-\partial V / \partial \qir = -b_3\qir^2$ to their respective
coordinates when the amplitude of the oscillations are large, and the
lattice will be rectified along these coordinates for the reasons
described in the previous paragraph.  From my first principles
calculations of the energy surface, I find that the coefficients of
these terms are such that the lattice is rectified along the direction
that reverses the electric polarization.

As the pump amplitude is increased, the effects of the abovementioned
nonlinearities start to become noticeable. The oscillations of the
$\qp$ mode start to show higher frequency components due to the
presence of various nonlinear terms. Further increase of the pump
amplitude takes the dynamics to a highly nonlinear regime. In this
regime, the frequency at which the $\qp$ mode oscillates also changes.
Interestingly, there are two different mechanisms that change the
effective frequency of the $\qp$ mode. First, a large amplitude
oscillation of the $\qp$ mode causes a change in the effective spring
constant due to the $\frac{1}{4}a_4\qp^4$ nonlinearity in the energy
potential. In the equation of motion, this nonlinearity acts to modify
the frequency of the $\qp$ mode with a term $-\partial V / \partial
\qp = -a_4\qp^3$. The effective frequency
$\omp^{\textrm{eff}}\to\omp^2 \left( 1 + a_4 \qp^2(t)/\omp^2 \right)$
changes because the time-averaged $\omp^{\textrm{eff}}$ has a value
different from $\omp$ when the $\qp$ amplitude is large. The
rectification of the lattice along the $\qp$ and $\qir$ coordinates
provides the second cause for the modification of the effective
frequency. For example, the rectification along the $\qp$ coordinate
changes the effective frequency to $\omp^{\textrm{eff}} \to
\omp^2\left(1 + a_3\qp(t)/\omp^2\right)$ due to the
$a_3\qp^3$ term in the energy potential. The $h\qp^2\qir$
term in the energy potential similarly changes the effective frequency
as $\omp^{\textrm{eff}} \to \omp^2\left(1 + 2h\qir(t)/\omp^2\right)$.
Fig.~\ref{fig:dyn}(b) shows the results of the numerical integration
of the equations of motion for a pump amplitude that rectifies the
lattice close to a point where the polarization switches. 
Near the polarization reversal, the slope of the potential for the
$\qp$ mode is less steep, and the $\qp$ mode oscillates at a smaller
frequency.
In this regime, the pumped $\qir$ mode is oscillating at a displaced
position with an amplitude that changes the two Ti-apical O bond
lengths by 0.4 and 0.7 \AA. These are large amplitude oscillations,
but they are comparable to the change in the Ti-apical O distance of
0.6 \AA\ when a polarization switch occurs.

Fig.~\ref{fig:dyn}(c) illustrates the case where the lattice moves to
a far distance in the negative direction along the $\qp$ coordinate,
and this signals that the polarization has been switched. The pump
strength used in this instance causes the $\qir$ mode to oscillate at
a displaced position with an amplitude that changes the two Ti-apical
O bond lengths by 0.5 and 0.8 \AA.  In this regime, the oscillations
about the displaced position exhibit a strong nonlinear behavior with
the presence of a wide range of frequency components. Nevertheless,
the frequency component that has the largest spectral weight stiffens
once the displacement along $\qp$ coordinate advances through the
point of polarization reversal, although the frequency is still
smaller than $\omp$.

I find that the displacement along the $\qp$ coordinate shows a sudden
jump when the externally pumped $\qir$ amplitude is continuously
increased. This is consistent with the behavior of the energy
potential discussed above where a large $a_3\qp^3$ term causes an
abrupt change in the position of the minimum of the $\qp$ mode as
$\qir$ is continuously increased. As a function of the pump amplitude,
the displacement along the $\qp$ coordinate continuously increases
from a value of zero to $\sim$$-$1.5 \AA
$\sqrt{\textrm{amu}}$. However, a further increase of the pump
amplitude causes the $\qp$ mode to oscillate about a displaced
position of $\sim$$-$9.0 \AA $\sqrt{\textrm{amu}}$. This indicates
that the electric polarization switches in an abrupt, discontinuous
manner when the $\qir$ mode is externally pumped. Such a behavior can
also be gleaned from the change in the frequency of the $\qp$ mode as
the polarization reversal happens. I find that the frequency of the
$\qp$ mode decreases by up to 60\% as it is displaced along this
coordinate. But it does not soften completely to zero as the
polarization switch occurs and the frequency starts to increase again.

In the study presented here, nonlinear couplings between two phonon
modes and their dynamics when the higher frequency mode is externally
pumped has been used to predict that ferroelectric materials can be
switched using mid-infrared pulses. However, in real materials there
are additional dynamical degrees of freedom, and this has two main
implications. First, scattering with other degrees of freedom will
cause the phonon modes to be damped. Therefore, the rectifying force
along the $\qp$ coordinate exists only as long as the $\qir$ mode is
being externally pumped. Second, other degrees of freedom also respond
to the displacement of the lattice along the $\qp$ coordinate. If the
pump pulse is long enough, other degrees of freedom relax relative to
the switched state, and this forms an energy barrier that prevents the
lattice to move back to the initial state even in the absence of the
pump.
A more detailed theoretical study based on molecular dynamics
simulations would be required to ascertain the time it would take to
form the energy barrier.

As mentioned above, I performed a full relaxation of the lattice
starting from a structure that corresponds to a $\qp$ displacement of
$-$8.0 \AA $\sqrt{\textrm{amu}}$ and found that the lattice indeed
relaxes to the symmetrically equivalent switched state. The relaxation
of the whole lattice to the symmetrically equivalent state with
reversed polarization provides a mechanism for repeated switching
because the lattice again experiences a unidirectional force in the
switching direction along the $\qp$ coordinate when the lattice is
excited anew by mid-infrared pulse.

In addition to PbTiO$_3$, I investigated the nonlinear couplings
between the lowest and highest frequency $A_1$ modes of BaTiO$_3$ and
LiNbO$_3$. I find a large $g\qp\qir^2$ coupling between the lowest
frequency $A_1$ mode ($\qp$) and the highest frequency $A_1$ mode
($\qir$) in these materials as well. The sign of the coupling is such
that the electric polarization of these materials could also be
switched by pumping the $\qir$ mode with a mid-infrared
pulse. Therefore, the method illustrated here seems applicable in
general to all perovskite transition-metal oxide ferroelectrics.

\section{Summary}
In summary, I have illustrated that the polarization of PbTiO$_3$ can
be switched by exciting the highest frequency infrared active $A_1$
phonon mode of this material with a mid-infrared pulse. A large
amplitude oscillation of this mode provides a unidirectional force
along the lowest frequency $A_1$ phonon mode coordinate due to a
nonlinear coupling of the type $g\qp\qir^2$.  A displacement of the
lattice along the $\qp$ coordinate changes the electric polarization
and can bring the system near the symmetrically equivalent switched
state.  From my first principles calculations, I find that sign of the
coupling is such that the oscillations of the $\qir$ mode displaces
the lattice along the $\qp$ coordinate in the switching direction. I
also find that the switching occurs discontinuously because of the
presence of a large $a_3\qp^3$ term in the energy
potential, which abruptly moves the minimum of the $\qp$ mode as the
$g\qp\qir^2$ term gets continuously larger. 

In addition to PbTiO$_3$, I find the presence of a similar
$g\qp\qir^2$ coupling and a large $a_3\qp^3$ anharmonicity
in BaTiO$_3$ and LiNbO$_3$, and this type of nonlinear coupling seems
to be universally present in perovskite transition-metal oxide
ferroelectrics. Therefore, a selective excitation of the $\qir$ mode
using a mid-infrared pulse can be a general method to switch the
polarization of perovskite transition-metal oxide ferroelectrics.

\section{Acknowledgments}
I am indebted to Yannis Laplace for valuable discussions. I also
acknowledge Antoine Georges, Roman Mankowsky, Srivats Rajasekaran, and
Andrea Cavalleri for helpful comments and discussions.

\end{document}